\documentclass[structabstract]{aa}  
\usepackage{graphicx}
\usepackage{txfonts}
\usepackage{natbib}
\begin{document}
   \title{Spatially resolved properties of the grand-design spiral galaxy UGC 9837: a case for high-redshift 2D observations}

\titlerunning{Spatially resolved properties of UGC 9837}


   \author{K. Viironen\inst{1,2}
          \and
          S. F. S\'anchez\inst{1}
          \and
          E. Marmol-Queralt\'o\inst{1,2}
          \and
          J. Iglesias-P\'aramo\inst{3,1}
          \and
          D. Mast\inst{1,3}
          \and
          R. A. Marino\inst{4,1}
          \and
          D. Crist\'obal-Hornillos\inst{2,3}
          \and
          A. Gil de Paz\inst{4}
          \and
          G. van de Ven\inst{5,6}
          \and
          J. Vilchez\inst{3}
          \and
          L. Wisotzki\inst{7}
          }

   	  \institute{Centro Astron\'omico Hispano Alem\'an, C/ Jes\'us
            Durb\'an Rem\'on 2-2, E-04004, Almeria, Spain\\
   	  \email{kerttu@cefca.es}
          \and
	  Centro de Estudios de F\'isica del Cosmos de Arag\'on (CEFCA), C/
          Pizarro 1, 3$^a$, E-41001 Teruel, Spain
          \and
          Instituto de Astrof\'isica de Andaluc\'ia (IAA-CSIC), Glorieta de la Astronom\'ia s/n, Granada E-18008, Spain
          \and
          Departamento de Astrof\'{i}sica y CC. de la Atm\'{o}sfera, Universidad Computense de Madrid, Madrid 28040, Spain
          \and
          Institute for Advanced Study (IAS), Einstein Drive, Princeton, New Jersey 08540, USA 
          \and
          Max Planck Institute for Astronomy, K\"onigstuhl 17, D-69117 Heidelberg, Germany
          \and
          Astrophysical Institute Potsdam, An der Sternwarte 16, D-14482, Postdam, Germany}
	 
   \date{Received September 15, 1996; accepted March 16, 1997}

 
  \abstract
   {We carry out a detailed 2D study of the ionised gas in the local universe galaxy UGC9837. In nearby galaxies, like the galaxy in question here, the spatial distribution of the physical properties can be studied in detail, providing benchmarks for galaxy formation models.}
   {Our aim is to derive detailed and spatially resolved physical properties of the ionised gas of UGC 9837. In addition, we derive an integrated spectrum of the galaxy and study how varying spatial coverage affects the derived integrated properties. We also study how the same properties would be seen if the galaxy was placed at a higher redshift and observed as part of one of the high-z surveys.}
   {UGC9837 was observed using the PMAS PPAK integral field unit. The spectra are reduced and calibrated and the stellar and ionised components separated. Using strong emission line ratios of the ionised gas, the source of ionisation, the dust extinction, the star formation rate, the electron density and the oxygen abundance derived from a total integrated spectrum, central integrated spectrum and individual fibre spectra are studied. Finally, the same properties are studied in a spectrum whose spatial resolution is degraded to simulate high-z observations.}
   {The spatial distribution of the ionised gas properties is consistent with inside-out growing scenario of galaxies. We also find that lack of spatial coverage would bias the results derived from the integrated spectrum leading, e.g., to an under-estimation of ionisation and over-estimation of metallicity, if only the centre of the galaxy was covered by the spectrum. Our simulation of high-z observations shows that part of the spatial information, such as dust and SFR distribution would be lost while shallower gradients in metallicity and ionisation strength would be detected.}
   {}

   \keywords{Galaxies: individual: UGC9837 -- Galaxies: abundances –- Galaxies: ISM — Galaxies: stellar content -- Techniques: spectroscopic –- Stars: formation}

   \maketitle
%

\section{Introduction}

The local universe galaxies are fundamental anchor points when studying the evolution of galaxies with cosmological time. Their apparent scale-lengths are small enough to derive relatively high resolution information on their physical properties with the current technology. This information, in turn, provides constraints that any theory of galaxy evolution need to meet. The Lambda-Cold Dark Matter paradigm ($\Lambda$CDM) is currently the most widely accepted model describing the formation of structures in the Universe. In case of the spiral galaxies, this model leads to the so-called inside-out formation scenario of galaxies \citep[see, e.g.][]{white91,mo98,kepner99}, i.e. the galaxy bulge forms first and is followed by the more slow process of disc formation. This scenario leads to various predictions about the galaxy structure such as negative age gradients, negative gradients of the gas-phase heavy element abundances, and decreasing amount of dust with increasing galactocentric distance. These predictions can be tested observationally.

Integral field spectroscopy provides a powerful tool for studying the distribution of physical properties in well-resolved galaxies \citep[eg.,][]{rosales-ortega10,sanchez11a}. Previous works have explored the use of different integral field spectrophotometers for a detailed study of nearby galaxies. In particular, the SAURON project \citep{dezeeuw02} and its extension Atlas3D \citep{cappellari11} are focused on the analysis of early-type galaxies and bulges of spirals at $z < 0.01$, to study their kinematics and stellar populations. Due to the distance of their objects ($D_L < 42$ Mpc) and the field-of-view (FOV) covered by the SAURON instrument ($33\arcsec \times 41\arcsec$), this study is mainly restricted to the central part of galaxies. The study of nearby spiral galaxies has been addressed by the VENGA survey using the VIRUS-P spectrograph \citep[32 nearby spiral galaxies,][]{blanc10} , the DiskMass Survey combining PPAK and the SparsePak spectrograph \citep[146 nearly face-on galaxies,][]{bershady10}, and the PINGS survey \citep[17 nearby disky galaxies,][]{rosales-ortega10}, using the PPAK IFU. In particular, the mosaicking designed for the PINGS survey allowed to map the H\,{\sc ii} regions along the whole extension of the galaxies and to explore the two-dimensional metallicity structure of disks. The galaxies of our sample \citep{esther11} were selected in size to fit in the FOV of the PPAK instrument, and therefore, to map their physical properties in their whole extension in just one pointing.

We present here a detailed study of ionised gas in one galaxy, UGC9837. This Local Universe ($z\sim0.009$) galaxy is positioned nearly face-on towards us, making it a good candidate for 2D galactic disk studies. The data were observed as part of the PMAS/PPAK, one of the worlds widest FOV IFU, survey of 48 nearby galaxies presented by \citet{esther11}. This survey was part of the preparatory phase of the CALIFA survey \citep{sanchez10,sanchez11b}, a much larger IFS survey comprising observations of ~600 nearby galaxies.

We use our data to derive for UGC9837 2-D distributions of: 1) local nebular reddening estimates based on the Balmer decrement; 2) oxygen abundance distributions based on a suite of strong line diagnostics incorporating reddening-corrected H$\alpha$, H$\beta$, [O\,{\sc ii}], [O\,{\sc iii}], [N\,{\sc ii}], and [S\,{\sc ii}] line ratios; 3) measurements of ionisation structure in H\,{\sc ii} regions and diffuse ionised gas using the well-known and most updated forbidden-line diagnostics in the oxygen and nitrogen lines; 4) the distribution of the surface star formation rate (SFR) traced by the H$\alpha$ line intensity. The derived spatial distribution of these properties allows us to test the inside-out growing scenario of galaxies.

In addition to the 2-D maps, we also study the integrated values of the above mentioned properties. The advantage of integral-field spectroscopy in this respect is that the observed spectra can be combined to produce an integrated spectrum of the object, using the IFU as a large aperture spectrograph. From this high signal-to-noise integrated spectrum the real average spectroscopic properties of the galaxy can be derived. This way no assumptions need to be made, in contrary to previous studies (based on long-slit spectroscopy, for example) attempting to describe the average spectroscopic properties of the galaxy based on individual spectra taken at different regions. This way the integrated properties of the galaxies in the Local Universe can be compared with the corresponding properties of the unresolved high-redshift objects. The most similar approach would be the spectrum derived by using drift-scanning technique \citep[e.g.][]{moustakas06a}. The advantage of using an IFU with respect to the drift-scan technique is that all the spectra are obtained simultaneously and that the IFU spectra allows a comparison between the integrated and the spatially resolved properties of the galaxy.

Finally, based on our Local universe data, we study how the spatial properties of ionised gas could be seen in observations at higher redshift, such as those carried out by SINS \citep{forster06} or MASSIV \citep{epinat09,queyrel09} surveys with VLT/SINFONI.

The article is structured in the following way: Sec. 2 presents the galaxy UGC9837; Sec 3 describes the observations and data reduction; in Sec. 4 the data analysis is presented; in Sec. 5 the results are discussed and in Sec. 6 the summary and conclusion are given. Throughout the paper we adopt a standard $\Lambda$CDM cosmology with H$_0 = 70.4$, $\Omega_m = 0.268$ and $\Omega_{\lambda} = 0.732$.

\section{UGC9837}

\begin{figure}
   \centering
   \includegraphics[width=\columnwidth]{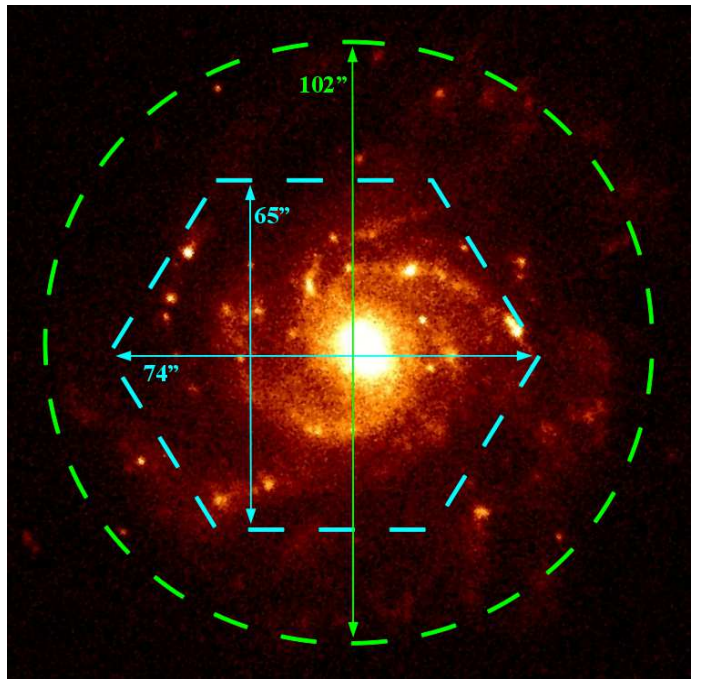}
      \caption{SDSS ($r$ band) image of UGC 9837. The cyan hexagon shows the approximate position of the PPAK pointing. The diameter of the green circle corresponds to the apparent diameter quoted in HyperLeda. North is up and East is left.}
         \label{fig:ugc9837}
   \end{figure}

\begin{table*}
\begin{minipage}[t]{\linewidth}
\caption{Photometric properties of UGC9837 as observed by different surveys at different pass-bands. All the fluxes are given in Janskys.}             
\label{tab:photom}      
\centering 
\renewcommand{\footnoterule}{}         
\begin{tabular}{|r r|r r|r r|r r|}     
\hline
\multicolumn{2}{|c|}{SDSS} & \multicolumn{2}{|c|}{2MASS} & \multicolumn{2}{|c|}{IRAS}& \multicolumn{2}{|c|}{GALEX}\\ 
\hline     
$u$ &2.83E-03 & $J$ &1.03E-02 & 12 $\mu$m &$<$ 5.82E-02 &FUV &2.73E-03\\
$g$ &7.23E-03 & $H$ &1.20E-02 & 25 $\mu$m &$<$ 6.05E-02 &NUV & 3.29E-03\\ 
$r$ &1.19E-02 & $K$ &9.37E-03 & 60 $\mu$m &2.83E-01 & &\\
$i$ &1.45E-02 & & &100 $\mu$m &8.97E-01 & &\\
$z$ &1.51E-02 & & & & & &\\
\hline                  
\end{tabular}
\end{minipage}
\end{table*}

UGC9837 is a spiral galaxy of magnitude 14.6 in the $g$ band, located at redshift $z\sim0.009$, and morphologically classified as SAB(s)c \citep{devaucouleurs91}. In addition to the optical (SDSS), it has previously been observed at other wavelengths: UV \citep[GALEX,][]{gildepaz07}, NIR \citep[2MASS,][]{skrutskie06} and \citep[IRAS,][]{beichman88}, see Table~\ref{tab:photom}. In addition, this galaxy was analysed as part of the DiskMass Survey \citep{bershady10}.

The distance to this galaxy has been estimated to be 41.2 Mpc \citep{tully88} with a distance modulus of 33.07 mag. The galaxy is extended \citep[$2.0\arcmin\times1.8\arcmin$,][]{nilson73} and nearly face-on which permits the spatial properties of the disk to be well studied. The total apparent corrected magnitude according to LEDA database is 13.65 mag, thus giving an absolute magnitude of M$_B$ = −19.42 mag. \citet{pohlen06} reported the existence of a bar of length $R \leq 9\arcsec$ ($ \sim 1.64$ kpc at the distance of the galaxy). These authors also found a break in the surface brightness profile at $R \sim 50\arcsec$ , which almost corresponds to the optical diameter, D$_{25} \simeq 2\arcmin$.
 
\section{Observations and data reduction}

UGC9837 was observed on 23rd June 2007. The night was not photometric and the extinction was high (A$_V >0.2$) and variable. During the night the seeing varied from 1 to 1.8$\arcsec$. The observation were carried out using
the 3.5 m telescope of the Calar Alto Observatory and the Potsdam Multi Aperture Spectrograph, PMAS \citep{roth05} in the PPAK mode \citep{verheijen05,kelz06}. 

Fig. 1 shows the SDSS r-band image of UGC 9837 with
the approximate position of the PPAK FOV over-imposed. As
it can be seen, a single PPAK exposure covers approximately
the whole area contained within the half the radius of the
25 mag arcsec$^{-2}$ isophote. The centre of the PPAK FOV is
slightly displaced from the optical centre of the galaxy in order
to include some interesting H\,{\sc ii} regions to the East of the centre.

The PPAK fibre
bundle consists of 331 science fibres of 2.7\arcsec~diameter each, covering a
total hexagonal field of view (FOV) of $74\arcsec \times 64\arcsec$ with a
filling factor of $\sim 60$\%. The sky is
sampled by 36 additional fibres, distributed in 6 bundles of 6 fibres each,
located following a circular distribution at about 90\arcsec~of the centre and
at the edges of the central hexagon. The sky-fibres are distributed among the
science ones in the pseudo-slit in order to have a good characterisation of
the sky. In addition there are 15 fibres used for calibration purposes.

The V300 grating was used, covering the wavelength range $\sim 3700-7100$\AA~, resulting in an effective spectral resolution of $\sim8$ \AA~(FWHM). Three dithered 600 s exposures of UGC9837 were taken following a pattern of $\Delta$RA= $\pm1.15\arcsec$ and $\Delta$Dec= $\pm0.78\arcsec$ in order to have a complete coverage of the science field of view and to increase the spatial resolution. An additional 120 sec exposures of the spectrophotometric standard Hz 4 was obtained in order to perform the flux calibration. Also bias frames, arc lamp exposures and sky exposures were taken for data reduction purposes.

Data reduction was carried out using R3D \citep{sanchez06}, in combination with E3D \citep{sanchez04}. The reduction consists of the standard steps for fibre-fed integral-field spectroscopy, including the bias subtraction, cosmic ray rejection, and extraction of the individual spectra by a Gaussian-suppression technique \citep{sanchez06} in order to minimise the effect of cross-talk between adjacent fibres. The extracted spectra were stored in a row-stacked-spectrum (RSS) file \citep{sanchez04} and wavelength calibrated. Differences in the fibre-to-fibre transmission throughput were then corrected by comparing the wavelength-calibrated RSS science frames with the corresponding ones derived from the twilight sky exposures. Finally, the sky was subtracted and the spectra were flux calibrated.

After reducing the three individual pointings, the corresponding RSS frames were combined into a single frame using standard procedures included in R3D \citep[see also][]{sanchez11a,esther11}. First, each position table was combined in a single one by co-adding to each pointing the differential offsets recorded by the telescope. PMAS is equipped with a precise guiding system that allows to perform relative offsets with an accuracy of 0.1$\arcsec$. Then, a 10$\arcsec$ radius aperture spectrum is extracted from the expected centre of the galaxy in each of the individual pointings, taking into account the offsets. These three spectra are compared, and if a difference larger than a 5\% is found, a rescaling is applied to match the intensity level of the first pointing. With this procedure possible variations in the transparency between the different exposures is compensated for.

The resulting data were then spatially re-sampled in a regular grid data-cube, by interpolating the information in each individual spatial pixel at each individual wavelength. This is done by using the interpolation routines included in E3D \citep{sanchez04}, by adopting a Natural-Neighbour non-linear interpolation scheme and a final pixel scale of 1$\arcsec$/pixel for the resulting data-cube. The Natural-Neighbour routine adopted here is based on the Voronoi tessellation of a discrete set of spatial points. For each particular pixel within the new grid the flux is assumed to be a weighted average of the n-adjacent ones:

\begin{displaymath}
G(x,y)=\displaystyle\sum\limits_{i=0}^n w_iF(x_i,y_i)
\end{displaymath}

\noindent where the n-neighbours are selected as the adjacent ones if the new pixel was included in the previous grid and a Voronoi tessellation was performed. The weights correspond to the fraction of area that the new tessellation will remove from the old one, if the new pixel is included. These weights are renormalised in order to conserve the integrated flux, by taking into account the number of adjacent pixels considered in each interpolation. It is important to note here that this interpolation does not require to fill gaps between the fibres, which would be the case if only one single pointing procedure was adopted. As already shown in \citet{sanchez07b}, where this procedure was presented for the first time, the dithering guarantees a complete coverage of the FOV, and the interpolation routine resamples the data, without creating new values in areas not covered by the IFU.

Finally, the flux calibration of the resulting datacube was adjusted using the SDSS photometry. This was done because even though the relative flux calibration of our data should be reliable, there are many uncertainties that can affect the goodness of the absolute spectrophotometric calibration. The photon-noise from the low surface brightness regions of the galaxy and the sky-background noise affect its quality. Also the sky subtraction entails inaccuracies. In addition, the night of the observations of UGC9837 was not photometric. Recalibrating our data based on the SDSS photometry available for UGC9837 is possible, because our IFS data allows us to carry our spectrophotometry over a wide enough aperture so that the comparison with a corresponding imaging photometry can be carried out without worrying about the PSF effects. 

\begin{figure}
   \centering
   \includegraphics[width=0.7\columnwidth,angle=-90]{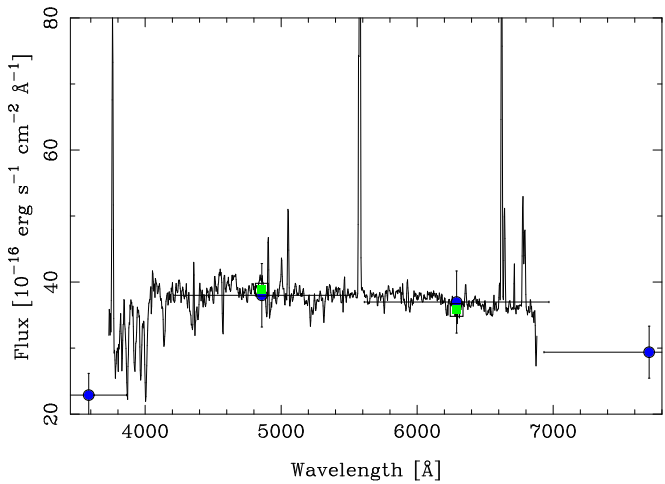}
      \caption{The flux re-calibration based on SDSS photometry. Our spectrum of the 30\arcsec~diameter central region of UGC9837, scaled by the average offset between it and the SDSS $g^\prime$ and $r^\prime$ band photometry, is shown as a black line. The SDSS $u^\prime, g^\prime, r^\prime$, and $i^\prime,$ band photometry are shown as blue dots and our spectrophotometry before re-scaling at $g^\prime$ and $r^\prime$ bands as green squares.}
         \label{fig:recalib}
   \end{figure}

Of the five SDSS filters our spectrum covers the pass-bands of two, namely the $g^\prime$ and $r^\prime$ filters. We measured the counts in these SDSS images inside a 30$\arcsec$ diameter aperture. Such aperture is large enough so as to avoid problems in the compared photometry regarding the exact position of the aperture centre and the exact radius of the aperture. These counts were converted to flux following the counts-to-magnitude prescription in SDSS documentation\footnote{http://www.sdss.org/dr7/algorithms/fluxcal.html\#counts2mag}. The corresponding spectrophotometry was then derived from our data, summing the flux of individual fibres inside a 30$\arcsec$ diameter circle and convolving this spectrum with the SDSS $g^\prime$ and $r^\prime$ filter pass-bands. Using these two data pairs, an average scaling factor between the SDSS photometry and our spectrophotometry was calculated. This is shown in Fig~\ref{fig:recalib}. The error in SDSS photometry is estimated to be 0.15 mag, including both the photon-noise and the background noise errors. The estimation is based on a statistical analysis in several different apertures of similar size on the back-ground subtracted $8\arcsec \times 8\arcsec$ post-stamp images. Our spectrum was then re-calibrated using the resulting average scaling factor (0.94), and the new calibrated spectrum was used in the following data analysis.

\begin{figure}
   \centering
   \includegraphics[width=\columnwidth]{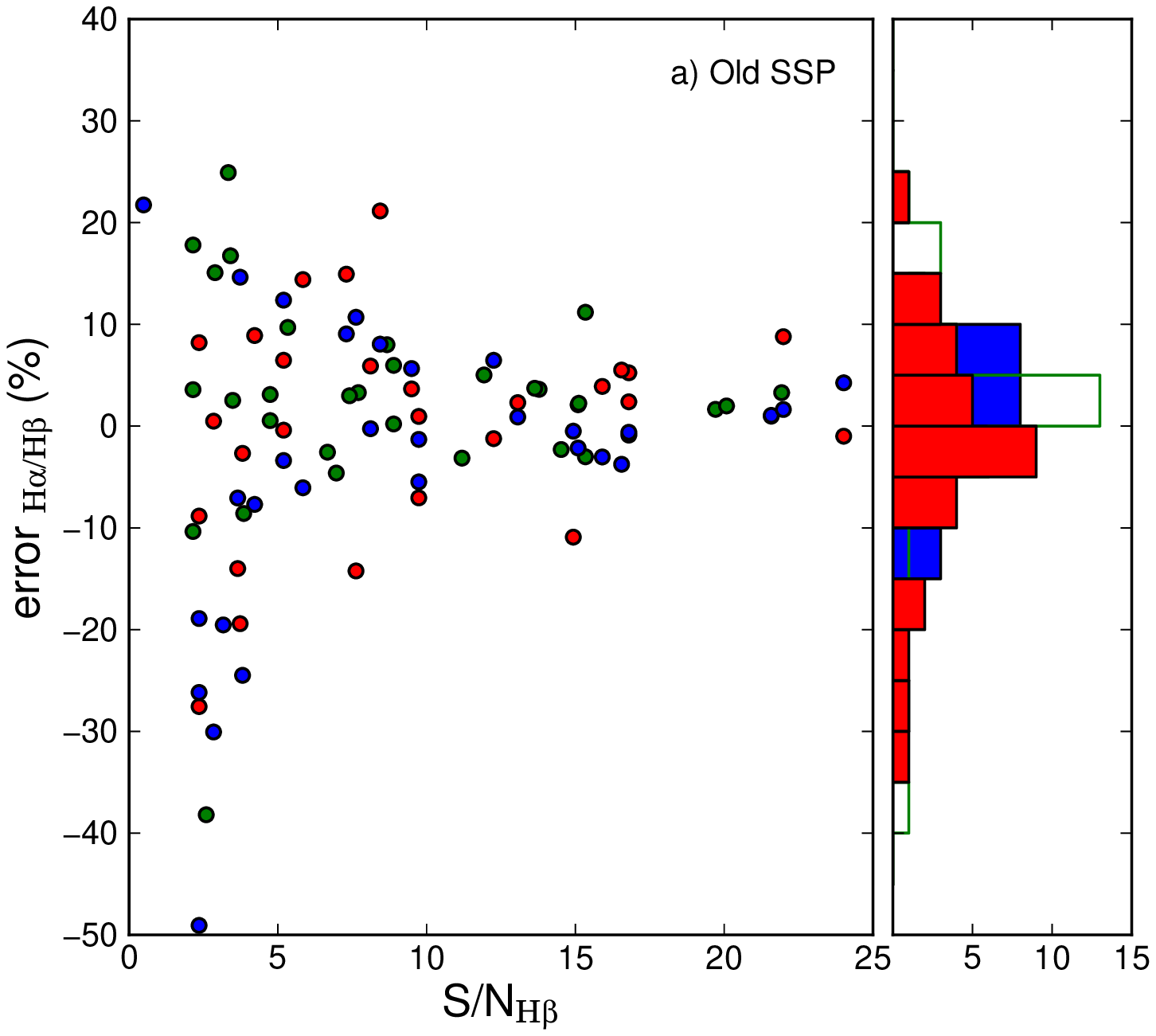}
   \includegraphics[width=\columnwidth]{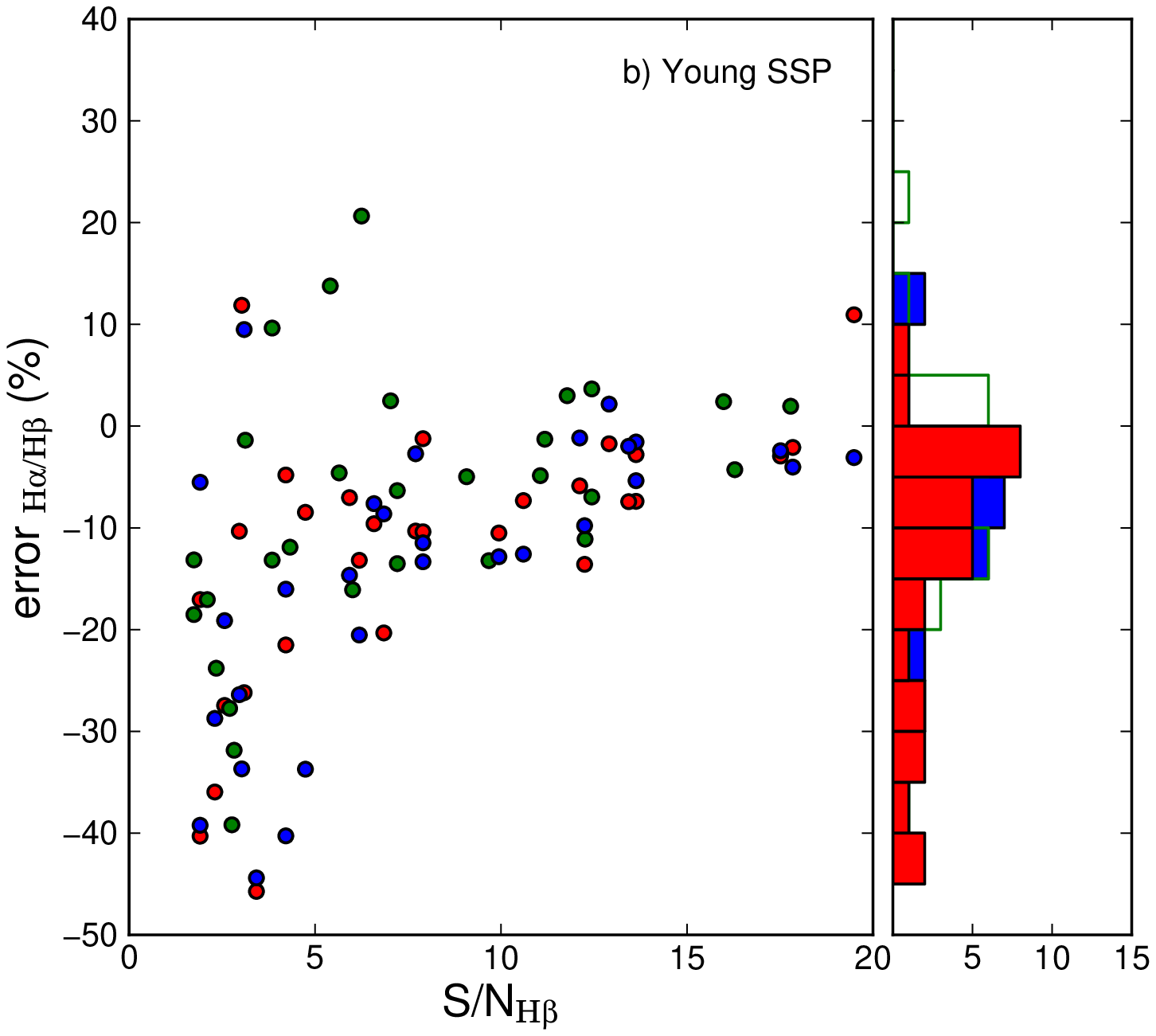}
      \caption{Error in the derived H$\alpha$/H$\beta$ ratio as a function of the S/N ratio of the H$\beta$ line for a young (1 Gyr, {\it Top}) and old (17.78 Gyr, {\it Bottom}) SSP with three different dust values, A$_V$ = 0.4 (red), 0.5 (green), and 0.6 (blue).}
         \label{fig:one2one}
   \end{figure}

\begin{figure}
   \centering
   \includegraphics[width=0.8\columnwidth,angle=-90]{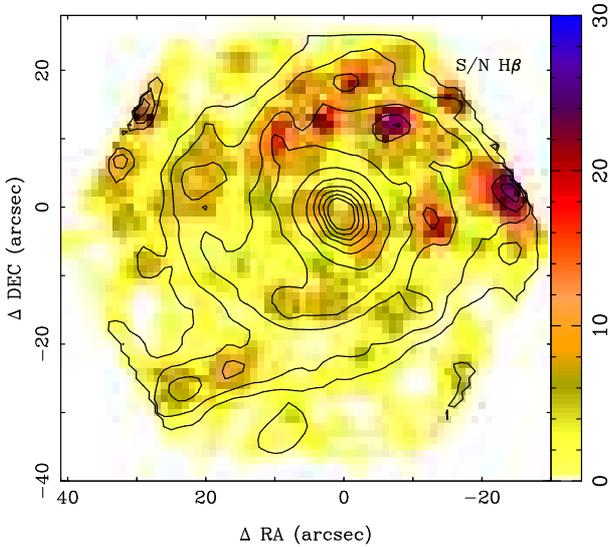}
      \caption{2-D map of the signal-to-noise ratio of the H$\beta$ line. Contours correspond to the V-band emission derived from the datacube at the flux levels: 0.025, 0.035, 0.053, 0.077, 0.105, 0.137, 0.172, 0.210, 0.251 and $0.295\times10^{−16}$ erg s$^{-1}$ cm$^{-2}$.}
         \label{fig:sn}
   \end{figure}

\begin{figure}
   \centering
   \includegraphics[width=\columnwidth]{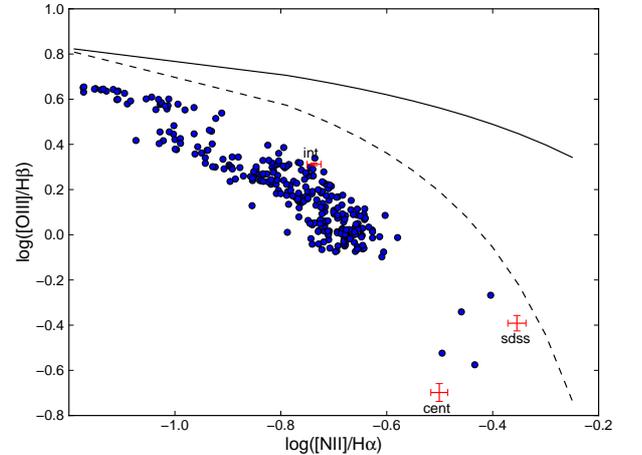}
      \caption{The N2O3 diagnostic diagram for our central spectrum (cent), the SDSS spectrum (sdss), the total integrated our spectrum (int), and the individual fibre-spectra of signal-to-noise ratio $> 5\sigma$ and H$\alpha$ line intensity $> 3 \times 10^{-16}$ erg s$^{-1}$ cm$^{-2}$ arcsec$^{-2}$ (blue dots). The solid and dashed lines separate the zones of starburst galaxies (below the lines) and AGNs (above the lines) as defined by \citet{kewley01} and \citet{kauffmann03}, respectively.}
\label{fig:n2o3int}
\end{figure}

\begin{figure}
   \centering
   \includegraphics[width=\columnwidth]{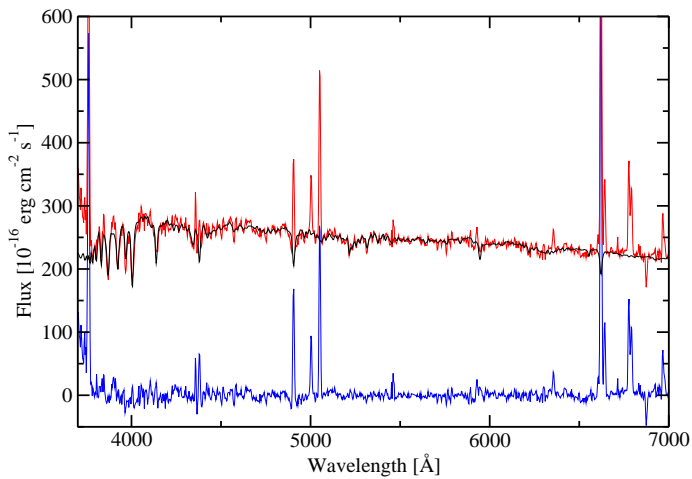}
      \caption{The integrated spectrum of UGC9837. The total integrated spectrum is shown in red, the best fitted combination of SSP models in black and the spectrum of the gas only in blue.}
         \label{fig:intspec}
   \end{figure}

\begin{table*}
\begin{minipage}[t]{\linewidth}
\caption{Measured and dereddened emission line fluxes (Flux, Flux\_dr, respectively) of the SDSS spectrum , and central and total integrated our spectra (SDSS, cent, tot, respectively) of UGC9837. All fluxes are first corrected for the underlying stellar absorption and then normalised to F(H$\beta$) = 1.0 and their errors are given in parenthesis.}             
\label{tab:intspec}      
\centering 
\renewcommand{\footnoterule}{}         
\begin{tabular}{l c|c c|c c|c c}     
\hline\hline       
Line ID & $\lambda$\footnote{Rest-frame} [\AA] &  Flux (SDSS) & Flux\_dr (SDSS) & Flux (cent) & Flux\_dr (cent) & Flux (tot) &  Flux\_dr (tot) \\
\hline
{}[O\,{\sc ii}] & 3726.8 & -- & -- & 5.91 (0.09) & 13.67 (0.88) & 3.42 (0.03) & 4.83 (0.10) \\ 
H$\gamma$ & 4340.5 & -- & -- & -- & -- & 0.43 (0.02) & 0.51 (0.02) \\ 
H$\beta$ & 4862.0 & 1.00 (0.06) & 1.00 (0.06) & 1.00 (0.06) & 1.00 (0.06) & 1.00 (0.02) & 1.00 (0.02) \\ 
{}[O\,{\sc iii}] & 4959.5 & 0.23 (0.02) & 0.22 (0.02) & 0.17 (0.02) & 0.15 (0.01) & 0.45 (0.01) & 0.44 (0.01) \\ 
{}[O\,{\sc iii}] & 5007.8 & 0.68 (0.05) & 0.65 (0.05) & 0.50 (0.04) & 0.45 (0.04) & 1.37 (0.02) & 1.31 (0.02) \\ 
{}[N\,{\sc ii}] & 6548.1 & 0.55 (0.02) & 0.42 (0.03) & 0.65 (0.02) & 0.30 (0.02) & 0.24 (0.01) & 0.17 (0.01) \\ 
H$\alpha$ & 6562.8 & 3.71 (0.08) & 2.86 (0.19) & 6.21 (0.09) & 2.86 (0.17) & 3.93 (0.02) & 2.86 (0.05) \\ 
{}[N\,{\sc ii}] & 6583.6 & 1.64 (0.06) & 1.26 (0.09) & 1.96 (0.07) & 0.90 (0.06) & 0.72 (0.02) & 0.52 (0.02) \\ 
{}[S\,{\sc ii}] & 6717.0 & 1.19 (0.07) & 0.90 (0.08) & 2.16 (0.07) & 0.95 (0.07) & 0.91 (0.02) & 0.65 (0.02) \\ 
{}[S\,{\sc ii}] & 6731.0 & 0.91 (0.05) & 0.69 (0.06) & 1.70 (0.07) & 0.74 (0.06) & 0.67 (0.02) & 0.48 (0.02) \\
\hline                  
\end{tabular}
\end{minipage}
\end{table*}

\begin{table*}
\begin{minipage}[t]{\linewidth}
\caption{Physical properties of UGC9837 derived from the SDSS spectrum, and central and total integrated our spectra (SDSS, cent, tot, respectively). }             
\label{tab:physpro}      
\centering 
\renewcommand{\footnoterule}{}         
\begin{tabular}{l c c c c}     
\hline\hline       
Property & SDSS & cent & tot\\
\hline
log([N\,{\sc ii}]/H$\alpha$) & -0.35$\pm$0.02 & -0.50$\pm$0.02 &  -0.74$\pm$0.01 \\
log([O\,{\sc iii}]/H$\beta$) & -0.39$\pm$0.03 & -0.70$\pm$0.04 & 0.312$\pm$0.006 \\
log(U) & -- & -4.21$\pm$0.04 & -3.473$\pm$0.009\\
c$_{\beta}$ & 0.38$\pm$0.09 & 1.13$\pm$0.08 & 0.46$\pm$0.03\\
A$_V^{gas}$ [mag]& 0.82$\pm$0.20 & 2.43$\pm$0.18 & 1.00$\pm$0.06 \\
12 + log(O/H) (O3N2, Pe 04) & 8.68$\pm$0.27 & 8.68$\pm$0.27 & 8.46$\pm$0.26\\
12 + log(O/H) (S2N2, Vi 07) & 8.42$\pm$0.21 & 8.25$\pm$0.19 & 8.13$\pm$0.17\\
12 + log(O/H) (Pi 10) & -- & 8.36$\pm$0.16 & 8.40$\pm$0.11\\
SFR [ M$_{\odot}$ yr$^{-1}$] & 0.06 & 0.13 & 2.48\\
\hline 
\hline                  
\end{tabular}
\end{minipage}
\end{table*}

\section{Data analysis}\label{analysis}

The first step to extract any physical information from the data set is to identify the emission lines and to decouple their emission from the underlying stellar population. After this, the emission line spectrum is analysed using standard equations for ionised gas.

\subsection{Deriving the emission line fluxes}\label{sec:fitting}

Population synthesis was used to model and subtract
the stellar continuum. This technique results in emission-line measurements
corrected (to a first-order) for stellar absorption. Many different techniques are now available to perform such a decoupling. The one applied here is the one used in the analysis of the PINGS sample \citep{rosales-ortega10} and the details
of this process are described in \citet{sanchez11a}. Shortly, the scheme followed to decouple the stellar population
and the emission lines is the following: i) A set of
emission lines is identified from the integrated spectrum. ii) For each spectrum in the data set, the underlying
stellar population is fitted by a linear combination of a grid
of Single Stellar Populations (SSP), after correcting for the appropriate systemic velocity and velocity dispersion (including the instrumental dispersion which dominates the total observed dispersion when the grating V300 is used), and taking into account the effects of dust attenuation. The models were created using the MILES templates \citep{vazdekis10} with three different ages (17.8, 1.0 and 0.09 Gyr) and two metallicities ($Z = 0.004$ and $0.03$). Prior to the linear fitting, the nebular and sky emission lines are masked. iii) The fitted
stellar population is subtracted from the original spectrum to get a residual
pure emission-line spectrum. iv) Finally, the intensities for each detected emission line are derived. We note that the adopted SSP template library has an impact on the derived properties of the SSP, and actually is not suitable for reliably deriving the SSP properties, but has little effect on the derived properties of the ionised gas \citep{esther11}.

Next, the individual emission-line fluxes were measured in each
spectrum by considering three spectral window regions: 1) [O\,{\sc iii}]3727\AA, 2) from H$\gamma$ to [O\,{\sc iii}]5007\AA, and 3) from [N\,{\sc ii}]6548\AA~to [S\,{\sc ii}]6731\AA. The windows were selected so that they cover all the nebular emission lines detected in the integrated spectrum. A simultaneous multi-component fitting was performed using
a single Gaussian function (for each emission line contained
within each window) plus a low order polynomial
(to describe the local continuum and to simplify the fitting
procedure) using FIT3D \citep{sanchez06b}. The central
redshifted wavelengths of the emission lines were fixed and
since the FWHM is dominated by the instrumental resolution,
the widths of all the lines were fixed as well. This procedure
decreases the number of free parameters and increases the
accuracy of the deblending process (when required). Line intensity fluxes were then measured by integrating the observed
intensity of each line. 

For the error analysis, an average continuum RMS of the original spectrum was calculated and added by bootstrapping method to the SSP+gas model spectrum provided by the fitting program. This was repeated 100 times and each time the fitting was done on the resulting spectrum. The sigma of the median absolute deviation of these 100 simulations was then assigned as the error in the measured emission/absorption line fluxes.

To test if, and how much, the decoupling procedure used affects the derived emission line fluxes, the following experiment was carried out. Different known emission line spectra (32 spectra derived from different parts of M74) were summed to a model spectra of both young (1 Gyr) and old (17.78 Gyr) SSP of \citet{vazdekis10}, assuming three values of dust, A$_V$ = 0.4, 0.5, and 0.6. The resulting spectra was then decoupled using the decoupling procedure introduced above, and the emission line fluxes measured. This was repeated 50 times adding random noise to the spectrum to get an idea about the involved errors. The derived emission line fluxes were then compared with the corresponding original fluxes. Fig. \ref{fig:one2one} shows the resulting difference (original - measured) in the measured H$\alpha$/H$\beta$ flux ratio. We see that the errors in the case of the young SSP are somewhat bigger as compared to the old SSP, but in both cases the errors are reasonable, $\leq 20$\% for S/N$>5$. In the case of the young SSP one is also biased to under-estimate the H$\beta$ emission flux due to underlying strong absorption. This bias can be expected to affect the extinction map derived in Sec.~\ref{sec:spat} from H$\alpha$/H$\beta$ ratio. The H$\alpha$ line map should not be strongly affected due to the strength of the H$\alpha$ emission, while the rest of the maps are not affected by the bias as line ratios are used in their derivation. We show in Fig.~\ref{fig:sn} the 2-D map of the signal to noise ratio of the H$\beta$ line, so that the reader can have an idea of the expected spatial distribution of the errors associated to the decoupling procedure of ionised gas and stellar populations. In the center of the galaxy and along the arms the S/N ratio is generally $>5$. 

\subsection{Methods}

The emission line fluxes derived above reflect the physical properties of the galaxy. Its dust extinction can be derived from the H$\alpha$/H$\beta$ line ratio. Throughout the paper, we have adopted the extinction law of \cite{cardelli89} with $R_V=3.1$. We assumed the Balmer decrement for case B recombination of ionised gas of temperature $10^4$ K and density of 100 cm$^{-3}$ \citep{osterbrock89}, i.e., I(H$\alpha$)/I(H$\beta$) $\sim 2.86$, as the unextincted expected value.

The nature of the ionisation is explored based on the log([N\,{\sc ii}]/H$\alpha$) vs. log([O\,{\sc iii}]/H$\beta$) (N2O3) diagnostic diagram of \citet{baldwin81} \citep[see also][]{veilleux87}. This diagnostic is insensitive to dust extinction, as the needed emission lines are very nearby in wavelength. This diagram allows to separate the ionisation dominated by star-formation from the one dominated by an Active Galactic Nucleus \citep[AGN, see e.g.,][]{kewley01,kauffmann03}.

The degree of the ionisation is traced by the ionisation parameter $U$. The lines available in our spectrum allow us to derive the ionisation parameter from the ratio [O\,{\sc ii}]/[O\,{\sc iii}]=$\lambda 3727/(\lambda4959+\lambda5007$) using the formula \citep{diaz00}:

\begin{displaymath}
\log U = -0.80\log (\mbox{[O\,{\sc ii}]/[O\,{\sc iii}]}) - 3.02.
\end{displaymath}

The electron density is derived from the [S\,{\sc ii}]6717\AA/6731\AA~ doublet ratio assuming a typical electron temperature for star forming regions, $T_e=10000$ K \citep[e.g.][]{osterbrock89}.

For oxygen abundance determination we use two empirical calibrations: O3N2=log\{([O\,{\sc iii}]$\lambda$5007/H$\beta$)/([N\,{\sc ii}]$\lambda$6583/H$\alpha$)\} indicator of \citet{pettini04} and the new calibration relations of \citet{pilyugin10} based on strong emission lines of oxygen, nitrogen and sulphur. The latter calibration relations are given separately to three classes of H\,{\sc ii} regions: cool, warm and hot, and the criteria how to classify the H\,{\sc ii} regions is given in \citet{pilyugin10}. Even though we are not studying individual H\,{\sc ii} regions here, we use the same criteria to decide which equation to use for deriving the oxygen abundance from each spectrum. Finally, we compare the results from the empirical calibrations to the ones derived from the model based log(H$\alpha$/[S\,{\sc ii}]$\lambda\lambda$6717+6731) vs. log(H$\alpha$/[N\,{\sc ii}]$\lambda$6583), S2N2, indicator of \citet{viironen07}. Instead of more commonly used methods, such as the R23 index \citep{kewley02}, we choose to use the O3N2 and S2N2 indicators because the SDSS spectrum with which we are comparing our data does not cover the [O\,{\sc ii}]$\lambda$3727 line.

Finally, the star formation rate (SFR) of the galaxy is studied based on its H$\alpha$ line intensity. The extinction corrected line fluxes are transformed to absolute luminosities and the corresponding SFRs are then derived adopting the classical relation between the SFR and the luminosity of the H$\alpha$ line \citep{kennicutt98}.

\section{Results}

In this section the results of the data analysis introduced above carried out for integrated and spatially resolved spectra of UGC9837 are presented.

\subsection{Integrated spectra}\label{sec:intspec}

To create an integrated spectrum, all the individual fibre-spectra were convolved with the Johnson $V$ filter pass-band and only those spectra with fluxes greater than $10^{-18}$ erg cm$^{-2}$ s$^{-1}$ were summed. This was done in order to not to take into account the data of the fibres that do not contain much signal or do not contain signal at all as the fibres are sampling the regions where the intrinsic flux from the galaxy is low or null (e.g. the borders of the galaxy, intra-arm regions). A similar approach was adopted in the analysis of the integrated spectrum of NGC628, presented by \citet{sanchez11a}, within the PINGS survey. We derive two integrated spectra, one covering only the central parts of the galaxy and another covering the whole data cube and compare the physical properties derived from these spectra. For comparison, we carry out similar analysis for the SDSS spectrum, which covers only a small central region of the galaxy.

To derive the central spectrum, the entire pixels inside a $6\arcsec$ diameter are summed. The diameter of the SDSS spectrum aperture is 3$\arcsec$. Since we are only interested in seeing quantitatively the differences in the physical properties derived in the centre as compared to those derived when (almost) the whole galaxy is covered, we did not try to exactly match the SDSS spectrum aperture because it would require interpolation (and the corresponding errors) as our pixel size is $1\arcsec$. In addition, we do not know the exact position of the SDSS aperture. 

The total integrated spectrum of UGC9837 was created summing the spectra of all the fibres (fulfilling the $V$-band flux criterion above). This spectrum, with the ionised and stellar components separated, is shown in Fig.~\ref{fig:intspec}.

Once the integrated spectra are created, the stellar and ionised components are separated and emission lines measured and analysed as described in Sec.~\ref{analysis}. Table \ref{tab:intspec} lists the emission lines and their fluxes with the corresponding errors in the SDSS spectrum, in our central spectrum, and in the total integrated spectrum. Table \ref{tab:physpro} lists the physical properties with their corresponding errors derived for UGC9837 from the three above mentioned spectra. The errors in the oxygen abundance include, in addition to the the total errors propagated from the errors in our data, the errors in the models themselves: 0.25 dex for the O3N2 method \citep{pettini04}, 0.075 dex for the new calibrations of \citet{pilyugin10}, and 0.15 dex as indicated by \citet{viironen07} for the S2N2 diagnostics. In addition, the N2O3 diagram is shown in Fig.~\ref{fig:n2o3int}. The derived densities for the total integrated spectrum, central our spectrum, and the SDSS spectrum, respectively, are N$_e$= 63$\pm$52, 144$\pm$125 and 107$\pm$173 cm$^{-3}$. The errors are large because these low densities are at the lower limit where the [S\,{\sc ii}] ratio is hardly sensitive to the density anymore.

A clear difference is encountered when comparing the location of the integrated total spectrum with the central spectra in the N2O3 diagram, the former pointing to higher excitation than the latter. This is not surprising considering that the total integrated spectrum includes the high excitation star forming regions of the spiral arms. Despite this difference, all the three spectra are located in the zone of star-formation in this diagram. Also the ionisation parameter $U$ indicates higher ionisation for the total spectrum as compared to the central one. Because the SDSS spectrum does not cover the [O\,{\sc ii}]$\lambda$3727 line, the degree of ionisation could not be derived from the SDSS spectrum. The extinction coefficient, $A_V$, derived from the total integrated spectrum is lower than the corresponding central value. The value derived from our central spectrum is about three times the value derived from the SDSS spectrum. In principle this could happen if in the SSP/ionised gas decoupling of our lower resolution data, the H$\beta$ absorption was underestimated. However, this would lead to a systematically higher values of all emission line fluxes (normalised to H$\beta$) in our spectrum as compared to the SDSS spectrum. This is not the case, as seen in Table \ref{tab:intspec}. Thus, we explain the difference in the two extinction values by short spatial scale variations of extinction, as the aperture sizes of the two spectra differ and we do not know the exact position of the SDSS spectrum as respect to our spectrum. In Sec.~\ref{sec:spat} we will show that there is significant spatial variations of extinction just near the centre of the galaxy.

There are clear shifts in the oxygen abundances derived using the different methods, the values derived from the S2N2 indicator being the smallest and the ones derived from the O3N2 indicator the highest. These biases probably derive from the fact that in the integrated spectrum, the fluxes of different lines can come from physically different zones of the galaxy, with different ionising clusters and metallicities. Also, the S2N2 method, the one giving the lowest abundances, is not yet calibrated in a consistent manner with the other bright-line methods. Finally, the presence of diffuse ionised gas (DIG) can modify the final line ratios. \citet{moustakas06b} showed that a strong DIG would increase both the N2 and S2 ratios, but less so the higher ionisation O3 ratio. As a result, both S2N2 and O3N2 indicators would slightly overestimate the metallicity. However, as the integrated spectrum is dominated by bright H\,{\sc ii} complexes, emitting strongly in O3, the effect of DIG in the abundance derived using O3N2 indicator is not necessarily significant \citep[see, for example][]{garcia-benito10}. In the new calibration relations of \citet{pilyugin10} the metallicity is calculated as 

\begin{eqnarray*}
12+\log(\mbox{O/H})=a_0+a_1\mbox{P}+&a_2&\log \mbox{O}3\\
+&a_3&\log(\mbox{N}2/\mbox{O}2)+a_4\log(\mbox{S}2/\mbox{O}2),
\end{eqnarray*}

\noindent where 

\begin{displaymath}
P=\frac{\mbox{O3}}{\mbox{O3}+\mbox{O2}} \;\;\;\;\mbox{and}\;\;\;\;  \mbox{O2}=\mbox{[O\,{\sc ii}]}\lambda\lambda3727+3729. 
\end{displaymath}

\noindent The effect of DIG on the relation is not discussed, but considering that its emission causes S2 and N2 to increase while O2, and especially O3, are less affected \citep{moustakas06b}, probably the presence of DIG leads to a slight overestimation of metallicity when this relation is used.

When comparing the total and central spectra, the oxygen abundances of the former derived from the O3N2 \citep{pettini04} and S2N2 \citep{viironen07} diagnostics lead to a lower oxygen abundance as compared to the corresponding values derived from the central spectra. This could be expected, as negative metallicity gradients are generally present in spiral galaxies. However, the abundances derived using the \citet{pilyugin10} calibration relations do not show such a trend, giving roughly the same result for the two spectra. In all cases the derived abundance is lower than the solar oxygen abundance \citep[12+log(O/H)=8.7,][]{scott09}.

We derive an integrated SFR of $\sim 2.4$ M$_{\odot}$/year. It is of the order of the values derived for spiral galaxies and  consistent with the picture of an unperturbed galaxy evolution \citep[eg.,][]{usui98}. As expected, being an additive property, the star-formation in the more reduced aperture of the central spectra is lower. When normalised by the aperture of each spectra, the SFR derived from the SDSS spectrum is actually the highest and then decreases when the aperture increases. This can be understood as in the larger aperture spectra the spectra of high star forming spiral arms is mixed with the spectra of intra-arm regions with low level diffuse H$\alpha$ emission.

\begin{figure*}
   \centering
   \includegraphics[width=\linewidth,angle=0]{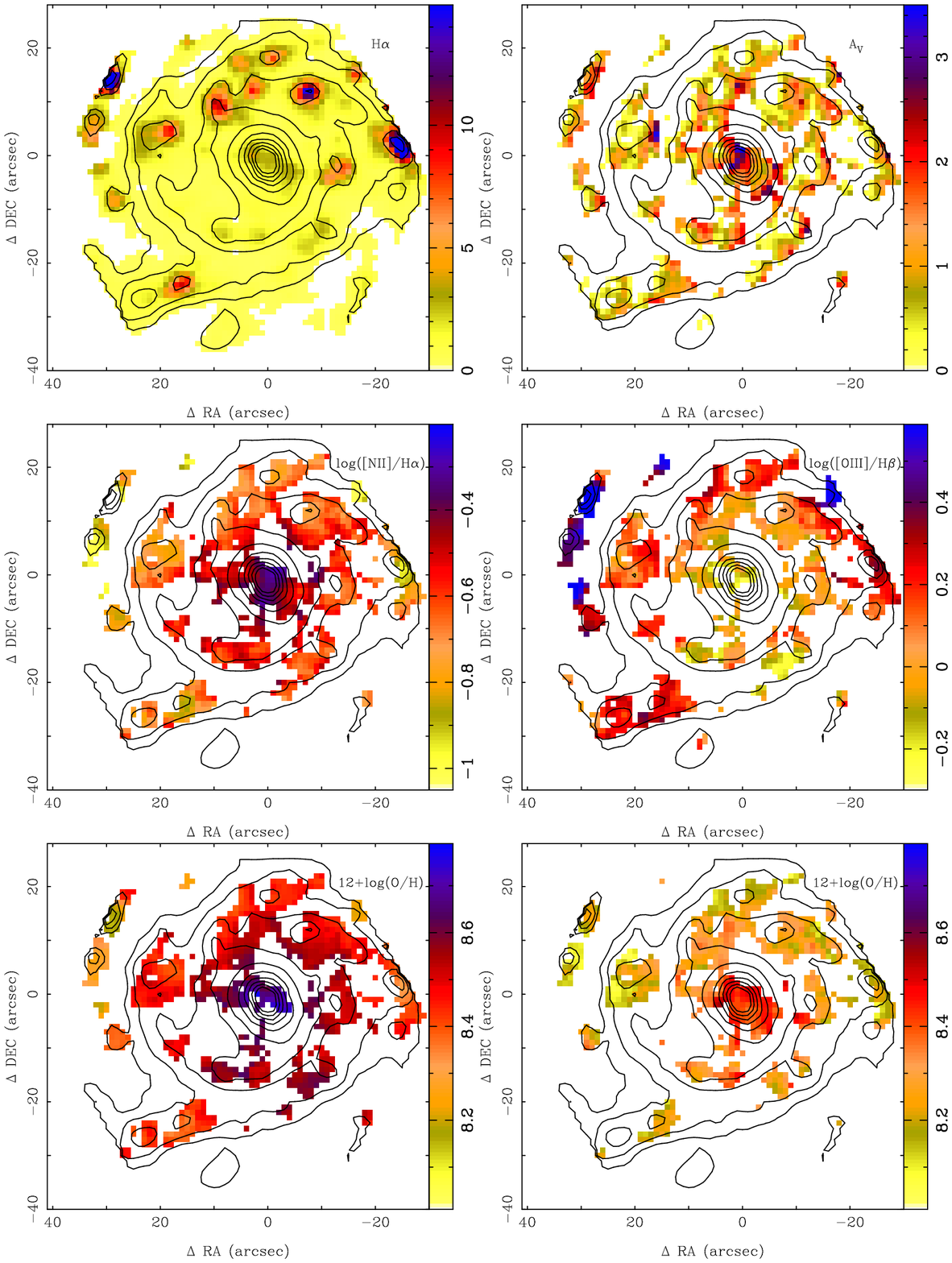}
         \caption{{\it Top Left:} H$\alpha$ line intensity map in units of $10^{-16}$ erg s$^{-1}$ cm$^{-2}$ arcsec$^{-2}$. {\it Top Right:} Dust extinction map of the ionised gas. {\it Middle left: } log([N\,{\sc ii}]/H$\alpha$) and {\it Middle right:} log([O\,{\sc iii}]/H$\beta$) diagnostic line-ratio maps. {\it Bottom Left:} Distribution of the oxygen abundance according to the O3N2 indicator of \citet{pettini04} and {\it Bottom Right:} S2N2 relation of \citet{viironen07}. The contours represent the $V$-band image derived from the data-cube and show the flux levels 0.025, 0.035, 0.053, 0.077, 0.105, 0.137, 0.172, 0.210, 0.251, and 0.295 $\times 10^{-16}$ erg s$^{-1}$ cm$^{-2}$.}
         \label{fig:maps1}
\end{figure*}

\subsection{Two-dimensional distribution of the gas properties}\label{sec:spat}

The real power of 2D spectroscopy resides in its capacity to map the physical properties of objects along their spatial extensions. We create the emission line maps for UGC 9837 from the final reduced datacube carrying out the fitting procedure described in Sec.~\ref{sec:fitting} for each individual spectra. For each map, the pixels with a flux $< 0.2\times 10^{-16}$ erg cm$^{-2}$
s$^{-1}$ at the corresponding wavelength are masked, in order to retain only the high signal-to-noise data
($\gtrsim 5\sigma$). When creating maps of emission line ratios, only those
locations are considered where all the involved lines are above this selected
minimum flux. Practically a map of any detected emission line, or a
combination of them, could be derived from the individual fibre spectra. Here
we present a number of maps corresponding to the most important emission lines
and derived quantities of interest.

As the dominant gas ionisation mechanism giving rise to the emission lines in this galaxy is due to hot (OB) stars, the H$\alpha$ line intensity map, shown in the top-left panel of Fig.~\ref{fig:maps1}, traces the star forming regions of the galaxy. The map shows a clear spiral structure with H\,{\sc ii} complexes of different sizes and morphology along the spiral arms. The brightest sources are located in the Northern arms. Also a clear presence of diffuse emission along the spiral arms and in the intra-arms region is visible.

\begin{figure}
   \centering
   \includegraphics[width=0.9\columnwidth,angle=-90]{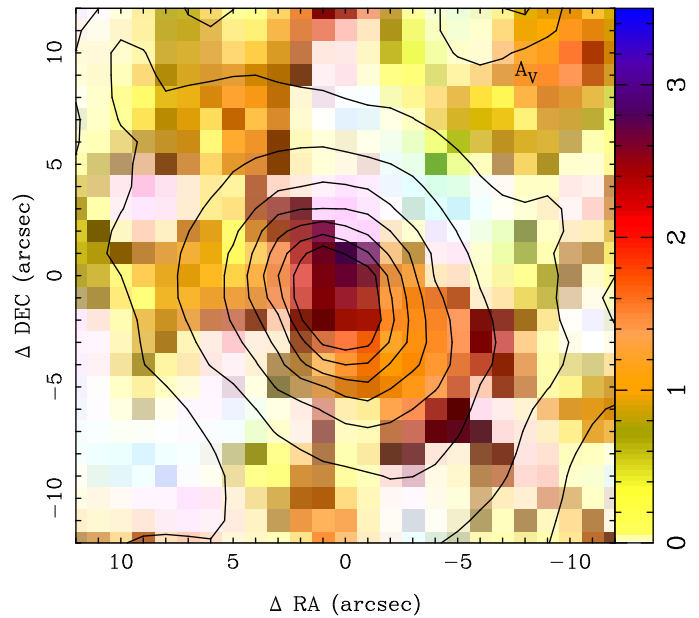}
         \caption{Closer look at the extinction variation (see Fig.~\ref{fig:maps1} Top Right) in the centre of the galaxy.}
         \label{fig:avin}
\end{figure}

The dust extinction was derived from H$\alpha$/H$\beta$ line ratio, as in the previous section, and is shown in Top right of Fig.~\ref{fig:maps1} in terms of $A_V$. The dust extinction is very variable, but as a general trend it is strongest in the bulge and then follows the spiral arms. We note that in the areas populated by young stars, i.e. expectedly in the spiral arms, the extinction may be overestimated by $\leq20$\% (see Figs.~\ref{fig:one2one} and \ref{fig:sn}.) A strong short-scale variation of extinction in the centre of the galaxy is visible (see also Figure~\ref{fig:avin}), as was already indicated in Sec.~\ref{sec:intspec} by the differences between the extinction values derived from our and SDSS central spectra. 


The dust extinction map described above is used to deredden the rest of the line maps and only the pixels covered by the dust map, in addition to the data on the lines in question, are considered. The middle panels of Fig. \ref{fig:maps1} shows the diagnostic line-ratio maps log([N\,{\sc ii}]/H$\alpha$), and log([O\,{\sc iii}]/H$\beta$) \citep{veilleux87}. The radial gradient is clear, the ionisation hardening towards the edge. This is indicated by the strengthening of the [O\,{\sc iii}]/H$\beta$ ratio towards the edge while [N\,{\sc ii}]/H$\alpha$ ratio becomes smaller. No clear trend along the spiral arms is visible. As shown in Fig. \ref{fig:n2o3int} the values of both ratios at any H$\alpha$ emitting location in the galaxy are consistent with ionisation produced by hot OB stars according to both \citet{kewley01} and \citet{kauffmann03}.

Next we studied the spatial distribution of the oxygen abundance as for the integrated spectra above using for brevity only the O3N2 indicator of \citet{pettini04} and S2N2 diagnostics of \citet{viironen07}. The resulting maps are shown in the bottom panels of Fig.~\ref{fig:maps1}. As the comparison of the central and integrated spectra above already hinted, in both maps the oxygen abundance is clearly strongest in the centre and decreases towards the edge, showing a gradient of $\sim -0.4$ dex, but O3N2 about 0.3 dex offset from S2N2.

\begin{figure*}
   \centering
   \includegraphics[width=\linewidth,angle=0]{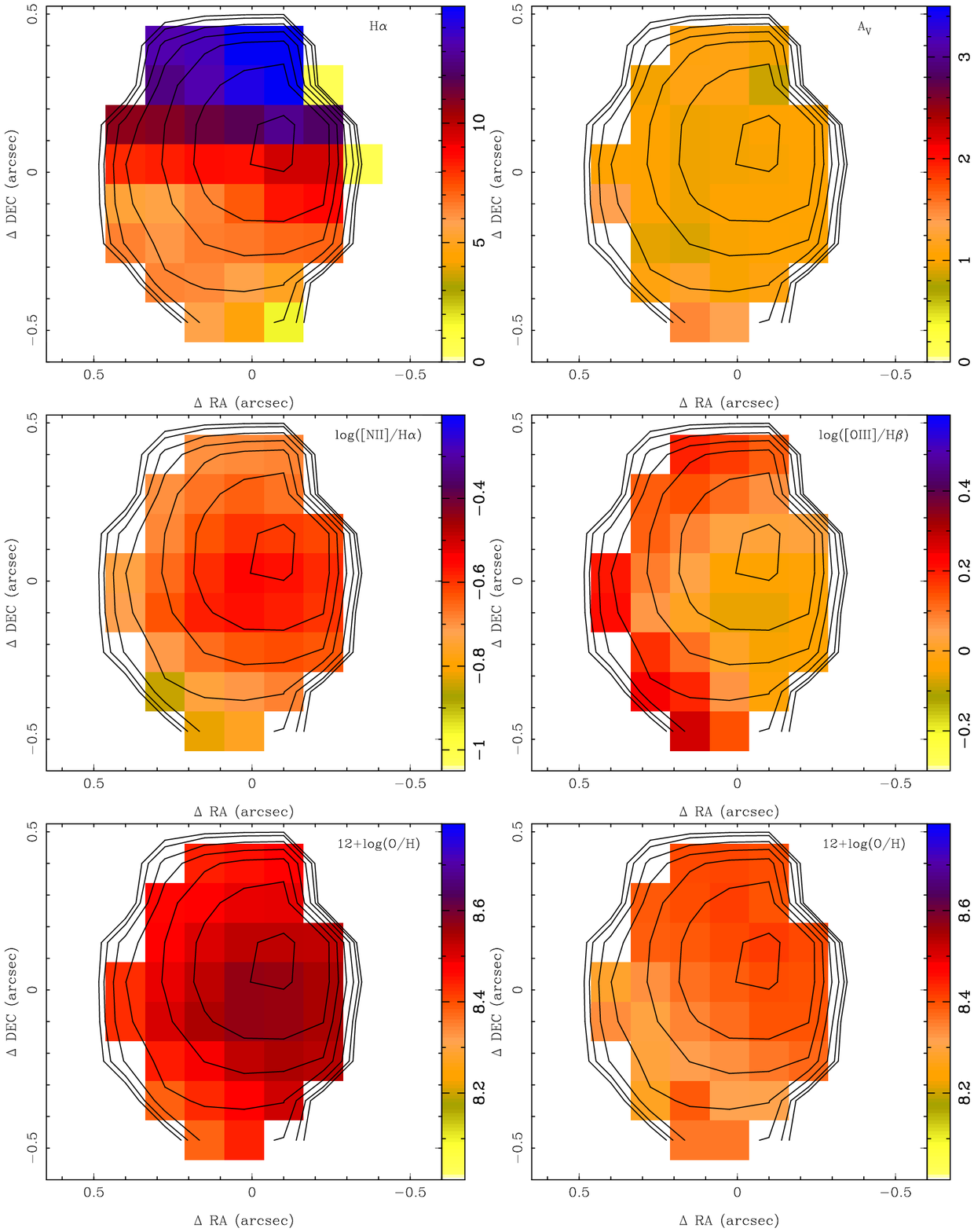}
         \caption{The same maps (but in units of $0.5\times 10^{-24}$ erg s$^{-1}$ cm$^{-2}$ arcsec$^{-2}$) as in Fig~\ref{fig:maps1} but as they would be seen by VLT/SINFONI if the UGC9837 was placed at redshift z=2.2.}
         \label{fig:maps2}
\end{figure*}

\subsection{Simulations of observations at higher redshifts}

In this section we study how the above maps would look like if the galaxy would be located at a higher redshift ($z\simeq1-2$) and observed as part of surveys such as SINS \citep{forster06} or ongoing MASSIV \citep{epinat09,queyrel09}, both using VLT/SINFONI. For this we have used a code adapted from the one used in \citet{eikenberry06}. The simulated spatial scale is $0.125\arcsec$/pix, and the data was depixelised accordingly, summing only entire pixels. Considering the original and simulated pixel sizes there is no differences in the depixelisation for simulated redshifts of 1.5 (typical for MASSIV) or 2.2 (typical for SINS). Next the flux was rescaled as

\begin{displaymath}
F=F_0\left(\frac{d_{orig}}{d_{simul}}\right) ^2\left(\frac{1+z_{orig}}{1+z_{simul}}\right),
\end{displaymath}

\noindent where $d_{orig}$ and $d_{simul}$ are the original and simulated luminosity distances, respectively, and  $z_{orig}$ and $z_{simul}$ the original and simulated redshifts. Finally, we assumed seeing conditions of $0.6\arcsec$ and convolved the data with two-dimensional Gaussian kernel of $0.6\arcsec$ FWHM. We accepted for the resulting datacube only the pixels where the convolution does not rely on the zero-padding used outside the image area. We have not added any noise to the simulated data as it depends on the observation strategy (e.g. integration time). However, in real observations at high redshift one would generally expect lower S/N ratios than in the lower-z observations. We neither considered the instrumental response of the simulated instrument at the wavelength where the line fluxes would be shifted at the simulated redshift nor we considered if the galaxy is intrinsically bright enough to be detected by the instrument at this redshift. Thus, to summarise, our simulation shows what kind of data could be obtained on a galaxy like UGC9837 if it was detected by an IFU of $0.125\arcsec$ pixel scale in typical observing conditions of $0.6\arcsec$ seeing. At the level of this simulation the picture would change only slightly if instrument like Keck/OSIRIS with $0.1\arcsec$ pixel scale was simulated while if a smaller pixel scale (which generally leads to a smaller FOV) was assumed, the result would be improved.

In Fig.~\ref{fig:maps2} we show the simulated maps (at z=2.2) derived from the maps in Fig.~\ref{fig:maps1}. As there would be no differences in the depixelisation and as we have not touched the S/N ratio, nor considered the instrumental response, the corresponding maps at z=1.5 would be equal but of higher flux. Comparing the maps in Figs.~\ref{fig:maps1} and \ref{fig:maps2} we see that in the high-redshift simulation the spatial information about the SFR (H$\alpha$ line map) is lost except for a gradient of stronger H$\alpha$ emission towards North, where the brightest H\,{\sc ii} regions are located. The information about the distribution of dust is lost completely. 

The combined effect of the latter two not only prevents us from deriving information about the SFR distribution along the galactic disc, but also lead to a bias in the level of SFR itself. As discussed in \ref{sec:spat}, the spatial distribution of the extinction by dust is very variable, being strongest in the centre and then following the spiral arms. The loss of this information leads to underestimation of absolute line fluxes, and SFR, in the dust rich parts of the galaxy, while opposite is true for the less dusty parts.

As for the relative line fluxes, the bluer line intensities are underestimated as respect to the redder lines in the dusty regions, especially in the centre and in the spiral arms. For this reason, we warn against using dust dependent indicators, such as emission line ratios of lines with large wavelength difference, or absolute line fluxes, if the distribution of dust extinction is not known.

The gradients in log([N\,{\sc ii}]/H$\alpha$) and log([O\,{\sc iii}]/H$\beta$) line ratio maps as well as in the 12 + log(O/H) can be still roughly appreciated in the simulated maps. However, the gradients appear shallower in the simulated maps as compared to the original ones. Shallower gradients can be expected as the spatial resolution of the simulated maps is worse so that zones of higher and lower ionisation and metallicity are mixed. This has its effect in the interpretation of the abundance gradients derived from observations with limited spatial resolution.

The inside-out growing scenario of galactic discs predicts the presence of negative metallicity gradients in spiral galaxies. However, different chemical evolution models  disagree on the time variation of these gradients. Some models predict that the gradients steepen with time \citep[e.g.][]{chiappini01} while other models predict the opposite \citep[e.g.][]{fo09}. Recently, \citet{queyrel11} published a study of metallicity gradients in a sample of galaxies at $z\sim 1.2$ based on SINFONI data. They found both negative, zero and positive gradients, explaining the latter by interactions. If a gradient was detected, it was in all cases shallow as compared to typical gradients of the nearby galaxies. Based on our simulation above, we argue that the gradients they measure are rather lower limits of the real abundance gradients of these galaxies.

\section{Summary and conclusions}

We have carried out a detailed study of the ionised gas of UGC9837 using the 2D data obtained with PMAS/PPAK \citep{esther11}. We have studied the integrated and spatially resolved properties of the ionised gas, deriving the nature of ionisation, ionisation parameter, oxygen abundance, SFR, and dust extinction ($A_V$) of the galaxy. First we compared the ionised gas properties derived from a spectrum covering only the centre of the galaxy (derived from our data and a SDSS spectrum) with the corresponding properties derived from a spectrum covering (almost) the whole galaxy. Clear differences were encountered as the spectra are sampling different zones of the galaxy. We conclude that deriving global properties of the galaxy from a spectrum of only partial spatial coverage would lead to significant biases. We derived for the galaxy a total SFR of $\sim 2.4$ M$_{\odot}$/year which is consistent with the picture of an unperturbed galaxy evolution for this object \citep[eg.,][]{usui98}. The integrated oxygen abundance derived using different strong line methods varied but all diagnostics lead to a subsolar abundance.

Next, the spatially resolved properties of the ionised gas were studied. We found that the ionisation becomes harder with increasing distance from the centre of the galaxy, while negative gradients of both oxygen abundance and dust extinction where uncovered. The $A_V$ distribution is consistent with tracing spiral arms, except for one (maybe two) arm(s), for which the dust distribution points to a leading nature of the arm(s). In general, our results are consistent with an inside-out growing scenario for disk galaxies, where the more evolved stellar populations and richer gas components with more dust, are located in the centre of the galaxies, and the less evolved stellar populations and metal poor gas, with less dust grains and stronger SFR and harder ionisation is located in the outer regions.

We also carried out a simulation how the galaxy would be seen if it was placed at a higher redshift and observed as part of surveys such as MASSIV or SINS. In our simulation the spatial information about the SFR and dust distribution is lost while radial gradients of ionisation strength and oxygen abundance can still be roughly distinguished. The gradients derived from the simulated data are shallower than the original (more realistic) gradients. These biases should be kept in mind when interpreting observations with limited spatial resolution. According to our Local Universe data, the dust distribution along the galactic disc can be very variable. If this information is lacking, dust independent indicators should be used for the study of the galaxy in question. The metallicity gradients derived from data with poor spatial resolution, in turn, are lower limits of the actual ones.

Finally, we want to remark the importance of a statistical sample of 2D data on nearby galaxies. The already existing data of the \citet{esther11} sample, and ultimately the data on a large sample of Local Universe galaxies being obtained by ongoing CALIFA \citep{sanchez10,sanchez11b} survey will allow analysis like the one presented here to be carried out in a statistical manner. This in turn will provide strong constraints on the theories of galaxy evolution and fix observational properties of galaxies in the Local Universe, which may have a potential impact in the interpretation of observed properties at higher redshifts.

\begin{acknowledgements}
      
\end{acknowledgements}

We thank the {\it Viabilidad, Dise\~no, Acceso y Mejora } funding program,
ICTS-2009-10, of the Spanish {\it Ministerio de Ciencia e Innovaci\'on}, for the
support given to this project.

This paper makes use of the Sloan Digital Sky Survey data. Funding for the
SDSS and SDSS-II has been provided by the Alfred P. Sloan Foundation, the
Participating Institutions, the National Science Foundation, the
U.S. Department of Energy, the National Aeronautics and Space Administration,
the Japanese Monbukagakusho, the Max Planck Society, and the Higher Education
Funding Council for England. The SDSS Web Site is http://www.sdss.org/. The SDSS is managed by the Astrophysical Research Consortium for the
    Participating Institutions. The Participating Institutions are the
    American Museum of Natural History, Astrophysical Institute Potsdam,
    University of Basel, University of Cambridge, Case Western Reserve
    University, University of Chicago, Drexel University, Fermilab, the
    Institute for Advanced Study, the Japan Participation Group, Johns Hopkins
    University, the Joint Institute for Nuclear Astrophysics, the Kavli
    Institute for Particle Astrophysics and Cosmology, the Korean Scientist
    Group, the Chinese Academy of Sciences (LAMOST), Los Alamos National
    Laboratory, the Max-Planck-Institute for Astronomy (MPIA), the
    Max-Planck-Institute for Astrophysics (MPA), New Mexico State University,
    Ohio State University, University of Pittsburgh, University of Portsmouth,
    Princeton University, the United States Naval Observatory, and the
    University of Washington.

Finally, we want to thank the anonymous referee for useful comments which helped us to considerably improve the article.

\bibliographystyle{aa}
\bibliography{oma}

\end{document}